\documentclass[aps,preprint,tightenlines,floatfix,groupedaddress,nofootinbib]{revtex4}

\usepackage{graphicx}
\usepackage{dcolumn}

\begin{document}
\bibliographystyle{h-physrev3}


\preprint{\vbox {\hbox{WSU--HEP--0501} \hbox{hep-ph/0503115}}}

\vspace*{2cm}

\title{\boldmath Higgs Sector of the Left-Right Model
with Explicit CP Violation}



\author{Ken Kiers}
\email{knkiers@tayloru.edu}
\author{Michael Assis}
\email{michael_assis@tayloru.edu}
\affiliation{Physics Department, Taylor University, 
236 West Reade Ave., Upland, Indiana 46989}

\author{Alexey A. Petrov}
\email{apetrov@physics.wayne.edu}
\affiliation{Department of Physics and Astronomy, Wayne State University,
Detroit, Michigan 48201}
\altaffiliation{}

\date{\today}

\begin{abstract}
We explore the Higgs sector of the Minimal Left-Right (LR) Model based on the 
gauge group $SU(2)_L\times SU(2)_R\times U(1)_{B-L}$ with explicit CP violation 
in the Higgs potential. Since flavour-changing neutral current
experiments and the small scale of neutrino masses both place
stringent constraints on the Higgs potential, we seek to determine whether 
minima of the Higgs potential exist that are consistent with current 
experimental bounds.  We focus on the case in which the right-handed symmetry-breaking 
scale is only ``moderately'' large, of order 15-50 TeV.  Unlike the case in which 
the Higgs potential is CP-invariant, the CP noninvariant case does yield viable 
scenarios, although these require a small amount of fine-tuning. We consider a 
LR model supplemented by an additional $U(1)$ horizontal symmetry, which 
results in a Higgs sector consistent with current experimental constraints and 
a realistic spectrum of neutrino masses.
\end{abstract}


\maketitle

\section{Introduction} \label{intro}

Left-Right (LR) models have long provided attractive extensions to the 
well-tested Standard Model (SM) of particle physics~\cite{patisalam1974,mohapatrapati,
mohapatrapati1975,senjanovic,mohapatrapaige,senjanovic1979,duka}. A typical LR model 
is constructed by enlarging the SM gauge group $SU(2)_L\times U(1)_Y$ to
the group $SU(2)_L\times SU(2)_R\times U(1)_{B-L}$ and thus contains
an extra set of $W$ and $Z$ gauge bosons compared to the Standard Model, as well
as an extended Higgs structure.  Minimal versions of the left-right model
usually contain one bidoublet Higgs boson, $\phi$, as well as left- and right-handed
triplet Higgs bosons, $\Delta_{L,R}$, although other constructions are also 
possible (see for example Ref.~\cite{Ma03}).

There are several reasons for the appeal of LR models. Among these is the fact 
that, unlike in the SM, the $U(1)$ symmetry of the theory has a physical interpretation
in terms of baryon and lepton number.  Also, the observed parity-odd nature 
of the low-energy weak interactions is seen to be the result of a spontaneously broken 
symmetry in the LR Model. Furthermore, it has been pointed out that, since parity is an exact 
symmetry of the theory at high energies, the strong CP problem has a natural
solution in the context of LR models~\cite{MoSen,BegTsao}.
In addition, the choice of a Higgs triplet to break left-right symmetry at
a high energy scale leads naturally to a seesaw relation
to explain the smallness of neutrino masses~\cite{mohapatra1980,Mohapatra:1980yp}.
Finally, the extended Higgs sector of the theory gives the possibility, at least in 
principle, of breaking CP spontaneously.

Many authors have focused on the possibility of spontaneous CP violation
in LR models, both in minimal versions as well as in extended versions 
containing more complicated Higgs sectors~\cite{branco,basecq,gunion,deshpande,gluza,bgnr}. 
The results in minimal versions of the LR Model have by and large been somewhat negative
regarding the possibility of generating spontaneous CP violation, while
investigations of more complicated models have had more encouraging results. 
A further complication in the case of the minimal model concerns the presence of 
neutral Higgs bosons having flavour-violating couplings to quarks (Flavour Changing 
Neutral Higgs (FCNH) bosons).  Recently, Barenboim et al studied the
decoupling limit of the minimal CP-invariant Higgs potential in the LR model~\cite{bgnr}.
They found that non-negligible CP violation in the vacuum state is generally 
accompanied by weak-scale non-standard Higgs bosons, which are ruled
out phenomenologically.  For CP-conserving vacuum states they found that it was
possible to produce an acceptable physical Higgs spectrum, but only at the cost of
extreme fine-tuning.

Another well-known complication in the Higgs sector of LR models
concerns the vacuum expectation value (VEV) of the neutral component of
the left-handed Higgs triplet, $\langle\delta_L^0\rangle\sim v_Le^{i\theta_L}$.
Assuming that all dimensionless coefficients in the Higgs
potential are of order unity, $v_L$ is related to its right-handed counterpart $v_R$
through a seesaw relation, $v_L\sim k^2/v_R$, where $k$ is a dimensionful quantity
of order of the weak scale.  The complication resides in the fact that, in general, 
the left-handed neutrino mass matrix contains a term that is proportional to $v_L$. 
Barring extreme fine-tunings, this term in the mass matrix forces $v_L$
to be of order a few eV or less, thereby forcing the right-handed scale, $v_R$, 
to be extremely large. To allow for the possibility of an observable right-handed
scale, many authors assume that $v_L=0$, a scenario that can be arranged by disallowing 
certain terms from the Higgs potential.  The offending terms in the Higgs potential have 
the form $\mbox{Tr}\!\left(\phi\Delta_R\phi^\dagger\Delta_L^\dagger\right)$ and may
be forbidden, for example, by imposing a symmetry under the discrete operation 
$\Delta_L\to\Delta_L$ and
$\Delta_R\to-\Delta_R$.  Unfortunately, invariance under this symmetry also disallows 
the Majorana Yukawa couplings that give rise to the seesaw relation for
neutrino masses~\cite{deshpande}.  Another approach is simply to assume that the
theory is embedded in a grand unified scheme that somehow disallows the troublesome terms
from the Higgs potential but still allows the desired Majorana Yukawa terms. This of course
does not solve this problem within the LR model framework.

In this paper we analyze the Higgs sector of the minimal LR Model in the case that CP 
is {\em not} a manifest symmetry of the unbroken theory, a scenario
that has not received much attention. In the most general case, the introduction of 
explicit CP violation corresponds to the introduction of only one complex coefficient
into the Higgs potential, i.e. only one extra degree of freedom. As we shall see below, the 
phase associated with this complex coefficient significantly alters the dynamics of the 
Higgs sector, and we find that the presence of CP violation in the vacuum
state no longer precludes the existence of an acceptable Higgs spectrum.  

In our calculation, in order to suppress $v_L$ to phenomenologically acceptable levels, 
we further assume the presence of an extra horizontal symmetry that is broken by a small 
parameter $\epsilon_H$~\cite{froggatt,nir1993,khasanov}.  An appropriate choice of 
charge assignments under this symmetry suppresses certain coefficients in the Higgs potential
and brings $v_L$ down to the eV scale for moderately light right-handed scales ($v_R\sim 15$~TeV, say).
The model still does contain a certain amount of fine-tuning.
In particular, one dimensionless
coefficient in
the Higgs potential needs to be ``unnaturally'' small (of order $k^2/v_R^2$).  
Since we restrict our attention to right-handed scales that are only ``moderately'' 
large, this fine-tuning is not as unnatural as 
it might be in general.

Our paper is organized as follows.  In Sec.~\ref{sec:model} we describe the model and 
perform the minimization of the Higgs potential to obtain relations among the various
Higgs VEVs.  Section~\ref{sec:mass_matrices} contains several approximate expressions
for the Higgs boson masses. Section~\ref{sec:numerical} contains our numerical results 
and in Sec.~\ref{sec:conclusions} we offer some concluding remarks.  The Appendix describes 
the derivations of the approximate expressions for the Higgs boson masses.

\section{The Model}\label{sec:model}

\subsection{Minimization of the Higgs Potential}

We consider a LR model with a single bidoublet field $\phi\sim(2,\overline{2},0)$
and triplet fields $\Delta_{L}\sim(3,1,2)$ and $\Delta_{R}\sim(1,3,2)$.
Under $SU(2)_L\times SU(2)_R$ the fermion fields in the theory transform
as $\Psi_{L,R}\to U_{L,R}\Psi_{L,R}$ and the Higgs fields transform as
$\phi\to U_L\phi U_R^\dagger$ and $\Delta_{L,R}\to U_{L,R}\Delta_{L,R}U_{L,R}^\dagger$.
We also define the field $\tilde{\phi}=\tau_2\phi^*\tau_2$, which
transforms in the same way as $\phi$.
The full Lagrangian of the theory (denoted below as LR Model) is invariant under the 
LR symmetry operation
\begin{eqnarray}
	\Psi_L\leftrightarrow\Psi_R,~~~~\phi\leftrightarrow\phi^\dagger,
	~~~~\tilde{\phi}\leftrightarrow\tilde{\phi}^\dagger,~~~~\Delta_L\leftrightarrow\Delta_R .
\end{eqnarray}
The most general LR symmetric Higgs potential that can be constructed from 
$\phi$, $\tilde{\phi}$, $\Delta_L$ and $\Delta_R$ may be parameterized as follows~\cite{deshpande},
\begin{eqnarray}
	V&=&-\mu_1^2\left[\mbox{Tr}\!\left(\phi^\dagger\phi\right)\right]
	-\mu_2^2\left[\mbox{Tr}\!\left(\tilde{\phi}\phi^\dagger\right)+
		\mbox{Tr}\!\left(\tilde{\phi}^\dagger\phi\right)\right]
	-\mu_3^2\left[\mbox{Tr}\!\left(\Delta_L\Delta_L^\dagger\right)+
		\mbox{Tr}\!\left(\Delta_R\Delta_R^\dagger\right)\right]\nonumber \\
	& & +\lambda_1\left[\mbox{Tr}\!\left(\phi\phi^\dagger\right)\right]^2
	+\lambda_2\left\{\left[\mbox{Tr}\!\left(\tilde{\phi}\phi^\dagger\right)\right]^2
		+\left[\mbox{Tr}\!\left(\tilde{\phi}^\dagger\phi\right)\right]^2\right\}
	+\lambda_3\left[\mbox{Tr}\!\left(\tilde{\phi}\phi^\dagger\right)
		\mbox{Tr}\!\left(\tilde{\phi}^\dagger\phi\right)\right]\nonumber\\
	& & +\lambda_4\left\{\mbox{Tr}\!\left(\phi\phi^\dagger\right)
		\left[\mbox{Tr}\!\left(\tilde{\phi}\phi^\dagger\right)
			+\mbox{Tr}\!\left(\tilde{\phi}^\dagger\phi\right)\right]\right\}
	+ \rho_1\left\{\left[\mbox{Tr}\!\left(\Delta_L\Delta_L^\dagger\right)\right]^2
		+\left[\mbox{Tr}\!\left(\Delta_R\Delta_R^\dagger\right)\right]^2\right\}\nonumber\\
	& & +\rho_2\left[\mbox{Tr}\!\left(\Delta_L\Delta_L\right)
			\mbox{Tr}\!\left(\Delta_L^\dagger\Delta_L^\dagger\right)
		+\mbox{Tr}\!\left(\Delta_R\Delta_R\right)
			\mbox{Tr}\!\left(\Delta_R^\dagger\Delta_R^\dagger\right)\right]
	+\rho_3\left[\mbox{Tr}\!\left(\Delta_L\Delta_L^\dagger\right)
		\mbox{Tr}\!\left(\Delta_R\Delta_R^\dagger\right)\right]\nonumber\\
	& & + \rho_4\left[\mbox{Tr}\!\left(\Delta_L\Delta_L\right)
			\mbox{Tr}\!\left(\Delta_R^\dagger\Delta_R^\dagger\right)
		+\mbox{Tr}\!\left(\Delta_L^\dagger\Delta_L^\dagger\right)
			\mbox{Tr}\!\left(\Delta_R\Delta_R\right)\right]\nonumber\\
	& &+\alpha_1\left\{\mbox{Tr}\!\left(\phi\phi^\dagger\right)\left[
		\mbox{Tr}\!\left(\Delta_L\Delta_L^\dagger\right)
		+\mbox{Tr}\!\left(\Delta_R\Delta_R^\dagger\right)\right]\right\} \nonumber\\
	& & +\alpha_2 \left\{e^{i\delta_2}\left[\mbox{Tr}\!\left(\phi\tilde{\phi}^\dagger\right)
			\mbox{Tr}\!\left(\Delta_R\Delta_R^\dagger\right)
		+\mbox{Tr}\!\left(\phi^\dagger\tilde{\phi}\right)
			\mbox{Tr}\!\left(\Delta_L\Delta_L^\dagger\right)\right]
		+\mbox{h.c.}\right\}\nonumber\\
	& & +\alpha_3\left[\mbox{Tr}\!\left(\phi\phi^\dagger\Delta_L\Delta_L^\dagger\right)
		+\mbox{Tr}\!\left(\phi^\dagger\phi\Delta_R\Delta_R^\dagger\right)\right]
	+\beta_1\left[\mbox{Tr}\!\left(\phi\Delta_R\phi^\dagger\Delta_L^\dagger\right)+
		\mbox{Tr}\!\left(\phi^\dagger\Delta_L\phi\Delta_R^\dagger\right)\right]\nonumber\\
	& & +\beta_2\left[\mbox{Tr}\!\left(\tilde{\phi}\Delta_R\phi^\dagger\Delta_L^\dagger\right)
		+\mbox{Tr}\!\left(\tilde{\phi}^\dagger\Delta_L\phi\Delta_R^\dagger\right)\right]
	+\beta_3\left[\mbox{Tr}\!\left(\phi\Delta_R\tilde{\phi}^\dagger\Delta_L^\dagger\right)
		+\mbox{Tr}\!\left(\phi^\dagger\Delta_L\tilde{\phi}\Delta_R^\dagger\right)\right] \; .
	\label{eq:Higgs_potl}	
\end{eqnarray}
All of the coefficients in the above expression are real, with the exception of 
the term proportional to $\alpha_2$, which has a phase $\delta_2$.   (
The coefficient $\alpha_2$ itself will be taken
to be real and non-negative in this work.)  Papers on this subject have nearly
universally set $\sin\delta_2$ to zero.  As pointed out in Ref.~\cite{bgnr}, however,
setting $\sin\delta_2$ to zero effectively means the vacuum state will not break CP.  
In contrast to this, we will see that it is quite natural to have large CP-violating phases
in the Higgs VEVs in the case
$\sin\delta_2\neq 0$.

We adopt the following notation for the various components of $\phi$, $\Delta_L$ and $\Delta_R$,
\begin{eqnarray}
	\phi=\left(\begin{array}{cc}
		\phi_1^0 & \phi_1^+ \\
		\phi_2^- & \phi_2^0 \\
		\end{array}\right) ,~~~
	\Delta_{L,R} = \left(\begin{array}{cc}
		\delta_{L,R}^+/\sqrt{2} & \delta_{L,R}^{++} \\
		\delta_{L,R}^{0} & -\delta_{L,R}^+/\sqrt{2}\\
		\end{array}\right).
	\label{eq:PhiDeltaLR}
\end{eqnarray}
The lepton Yukawa couplings that
are consistent with the gauge and parity
symmetries discussed above are
\begin{eqnarray}
	-{\mathcal L}_\textrm{\scriptsize Yukawa} = 
		\overline{\psi}_{iL}^\prime\left(
		G_{ij} \phi + H_{ij} 
		\widetilde{\phi}\right)\psi_{jR}^\prime +
		\frac{i}{2}F_{ij}\left(\psi^{\prime T}_{iL} C \tau_2
		\Delta_L\psi^{\prime}_{jL} +
		\psi^{\prime T}_{iR} C \tau_2
		\Delta_R\psi^{\prime}_{jR}\right)
		+ \textrm{h.c.} \; ,
	\label{eq:yuk}
\end{eqnarray}
where $F$, $G$ and $H$ are $3\times 3$ Yukawa matrices,
$C=i\gamma^2\gamma^0$ is the charge conjugation matrix
and primes denote gauge eigenstates.  
Upon spontaneous symmetry breaking, the neutral components of the Higgs boson fields obtain VEVs,
\begin{eqnarray}
	\langle \phi\rangle =\left(\begin{array}{cc}
		k_1/\sqrt{2} & 0 \\
		0& k_2e^{i\alpha}/\sqrt{2} \\
		\end{array}\right) ,~~~
	\langle \Delta_L \rangle= \left(\begin{array}{cc}
		0 & 0 \\
		v_Le^{i\theta_L}/\sqrt{2} & 0\\
		\end{array}\right),~~~
	\langle \Delta_R \rangle= \left(\begin{array}{cc}
		0 & 0 \\
		v_R/\sqrt{2} & 0\\
		\end{array}\right),
	\label{eq:HiggsVEVs}
\end{eqnarray}
where $k_1$, $k_2$, $v_L$ and $v_R$ all refer to the magnitudes
of the respective quantities.  Phenomenological considerations lead
to the conclusion that $v_L\ll k_1, k_2 \ll v_R$.  Furthermore, $k_1$ and
$k_2$ are at the weak scale, $k_1^2+k_2^2\simeq (246$~GeV$)^2$.
Gauge rotations have been used in the above expressions to eliminate
possible phases in the $k_1$ and $v_R$ terms~\cite{deshpande}.
The four complex neutral fields may then be expanded in terms of eight real fields as follows,
\begin{eqnarray}
	\phi_1^0 &=& (\phi_1^{0r} + i\phi_1^{0i}+k_1)/\sqrt{2},\label{eq:phi10} \\
	\phi_2^0 &=& (\phi_2^{0r} + i\phi_2^{0i}+k_2)e^{i\alpha}/\sqrt{2},\\
	\delta_L^0&=&(\delta_L^{0r} + i\delta_L^{0i}+v_L)e^{i\theta_L}/\sqrt{2},\\
	\delta_R^0&=&(\delta_R^{0r} + i\delta_R^{0i}+v_R)/\sqrt{2}. \label{eq:deltaR0}
\end{eqnarray}

Minimizing the potential in Eq. (\ref{eq:Higgs_potl}) with respect to $\phi_1^{0r}$,
$\phi_2^{0r}$, $\phi_2^{0i}$, $\delta_R^{0r}$, $\delta_L^{0r}$ and $\delta_L^{0i}$ 
leads to six equations that need to be satisfied by the coefficients of the potential.
By manipulating these expressions we obtain expressions for 
the $\mu_i^2$,
\begin{eqnarray}
	\frac{\mu_1^2}{v_R^2} & = &\frac{\alpha_1}{2}\left(1+
		\frac{v_L^2}{v_R^2}\right)-
		\frac{\alpha_3\xi^2}{2(1-\xi^2)}\left(1+
		\frac{v_L^2}{v_R^2}\right)+
		\left[\lambda_1\left(1+\xi^2\right)
			+ 2\lambda_4\xi\cos\alpha\right]\epsilon^2
			\nonumber\\
		& & + \left[\beta_2\cos\theta_L -
		\beta_3\xi^2\cos(\theta_L-2\alpha)\right]
			\frac{v_L/v_R}{1-\xi^2} \; ,
		\label{eq:mu1full}\\
	& & \nonumber \\
	\frac{\mu_2^2}{v_R^2} & = & \frac{\alpha_2}{2\cos\alpha}\left[
		\cos(\alpha+\delta_2)+
		\cos(\alpha-\delta_2)\frac{v_L^2}{v_R^2}\right]+
		\frac{\alpha_3\xi}{4(1-\xi^2)\cos\alpha}\left(
		1+\frac{v_L^2}{v_R^2}\right) \nonumber\\
		& & +
		\left[2\lambda_2\xi\cos(2\alpha)+
		\lambda_3\xi+
		\frac{1}{2}\lambda_4\left(1+\xi^2\right)\cos\alpha
			\right]\frac{\epsilon^2}{\cos\alpha}\nonumber\\
		& & +
		\left[\beta_1(1-\xi^2)\cos(\theta_L-\alpha)-
		2\beta_2\xi\cos\theta_L+
		2\beta_3\xi\cos(\theta_L-2\alpha)
			\right]\frac{v_L/v_R}{4(1-\xi^2)\cos\alpha}\; ,
		\label{eq:mu2full}\\
	& & \nonumber \\
	\frac{\mu_3^2}{v_R^2} & = & \rho_1
			\left(1+\frac{v_L^2}{v_R^2}\right)+
		\frac{1}{2}\left[\alpha_1\left(1+\xi^2\right)+
			\alpha_3\xi^2\right]\epsilon^2 \nonumber\\
		& & +
		2\alpha_2\left[
		\cos(\alpha+\delta_2)-
		\cos(\alpha-\delta_2)\frac{v_L^2}{v_R^2}\right]
		\frac{\xi\epsilon^2}{1-v_L^2/v_R^2}\; ,
		\label{eq:mu3full}
\end{eqnarray}
where we have defined $\xi\equiv k_2/k_1$ and $\epsilon\equiv k_1/v_R$.
Both $\xi$ and $\epsilon$ are small quantities in our analysis.
We also obtain three other equalities that must be satisfied,
\begin{eqnarray}
	& & \hspace{-1in}
	\left[(2\rho_1-\rho_3)-\frac{8\alpha_2\xi\epsilon^2
			\sin\alpha\sin\delta_2}
		{1-v_L^2/v_R^2}\right]\frac{v_L}{v_R} \nonumber\\
	& = & \left[\beta_1\xi\cos(\theta_L-\alpha)+\beta_2\cos\theta_L
		+\beta_3\xi^2\cos(\theta_L-2\alpha)\right]\epsilon^2 \; ,
		\label{eq:A5}\\
	& & \nonumber \\
	0 & = & \beta_1\xi\sin(\theta_L-\alpha)+\beta_2\sin\theta_L
		+\beta_3\xi^2\sin(\theta_L-2\alpha) \; ,
		\label{eq:A6}\\
	& & \nonumber \\
	& & \hspace{-1in}
	2\alpha_2\left(1-\xi^2\right)\left(1-v_L^2/v_R^2\right)\sin\delta_2
		\nonumber\\
	& = & \left\{2\xi\sin(\theta_L-\alpha)(\beta_2+\beta_3) + 
		\left[\sin\theta_L+\xi^2\sin(\theta_L-2\alpha)\right]\beta_1
			\right\}\frac{v_L}{v_R} \nonumber\\
	& & +\xi\sin\alpha\left[\alpha_3\left(1+v_L^2/v_R^2\right)+
			(4\lambda_3-8\lambda_2)\left(1-\xi^2\right)
				\epsilon^2\right]\; .
		\label{eq:A7}
\end{eqnarray}
Taking $\delta_2\to 0$ in Eqs.~(\ref{eq:A5})-(\ref{eq:A7}) gives
Eqs.~(A5)-(A7) in Ref.~\cite{deshpande}.
Equation~(\ref{eq:A5}) yields the well-known seesaw relation
for $v_L$, $v_L\sim(\mbox{some couplings})\times\epsilon^2 v_R$.  
Interestingly, $\rho_2$ and $\rho_4$ do not appear in the
above expressions and are thus not constrained by the first-derivative conditions.
(The first of these does, however, appear in an approximate expression for one of the 
Higgs boson masses and is thus constrained through a second-derivative condition.)

A very important feature of the above equations is that the left-hand side of Eq.~(\ref{eq:A7})
is generically non-zero.  This gives us considerably more freedom when solving the minimization
conditions compared to the situation considered throughout much of Ref.~\cite{bgnr}
(in which it was mostly assumed that $\sin\delta_2=0$).  In particular, in
our case we can easily obtain $\alpha_3={\cal O}(1)$ instead of ${\cal O}(\epsilon^2)$.  
This observation is crucial, since the neutral Higgs bosons with flavour non-diagonal couplings to
quarks have masses $\sim\sqrt{\alpha_3v_R^2/2}$ (see Eq.~(\ref{eq:malpha3_0}) below).
If $\alpha_3={\cal O}(1)$, these FCNH bosons attain masses at the scale $v_R$ and
are phenomenologically viable.  If, however, $\alpha_3$ is of order $\epsilon^2$, then the
FCNH bosons have masses at the weak scale, a scenario that presents considerable phenomenological
problems.  Let us examine this argument a bit more quantitatively.
Suppose for the moment that $\sin\delta_2=0$ and 
$v_L/v_R\ll 1$.  Suppose furthermore that $\sin\alpha \neq 0$, so that the vacuum
state breaks CP.  In order to satisfy Eq.~(\ref{eq:A7}) in this case, one must have
$\alpha_3\simeq (8\lambda_2-4\lambda_3)(1-\xi^2)\epsilon^2$; i.e., unless $\lambda_{2}$
is very large, $\alpha_3$ must be of order $\epsilon^2$, leading
to phenomenological difficulties.  If $\sin\delta_2\neq 0$, however,
things are quite different.  In this case Eq.~(\ref{eq:A7}) becomes
\begin{eqnarray}
	2\alpha_2\sin\delta_2 \simeq 
		\alpha_3\xi\sin\alpha \; 
	\label{eq:A7approx}
\end{eqnarray}
(where we have assumed that $v_L/v_R$ is negligibly small and that the various
Higgs coefficients are not anomalously large).  It is possible to 
satisfy this expression with $\alpha_3={\cal O}(1)$.  In particular, we can use
the above expression to determine
the phase of $\langle \phi_2^0\rangle$,
\begin{eqnarray}
	\alpha \simeq \sin^{-1}\left(\frac{2\alpha_2\sin\delta_2}
		{\alpha_3\xi}\right) .
	\label{eq:alphaapprox}
\end{eqnarray}
With the choice we make below for the charges under the approximate horizontal symmetry,
we have $\alpha_3\sim{\cal O}(1)$ and
$\alpha_2\sim{\cal O}(\xi)$.  In this case $\alpha$ is of order unity if $\delta_2$ is.
To summarize, if $\sin\delta_2$ is non-zero, one can in fact have a vacuum state that
violates CP while simultaneously evading the FCNH problem.  This possibility
was noted in Ref.~\cite{bgnr}, but was not explored in detail.

\subsection{A Broken Horizontal Symmetry}

Let us return to the seesaw relation for the VEVs derived from Eq.~(\ref{eq:A5}) and noted
above.  If all the coefficients in the Higgs potential are of
order unity one has $v_L\sim(\mbox{some couplings})\times\epsilon^2 v_R$ 
(note that $\epsilon^2v_R=k_1^2/v_R)$.  
If $k_1$ is of order the weak scale
and $v_R$ is of order 10's of TeV, then $v_L$ is of order a few GeV.  Such a value for $v_L$ is 
extraordinarily large from the point of view of neutrino physics, an observation that can be 
understood by examining the $3\times 3$ mass matrix for the light, mostly left-handed neutrinos.  
This mass matrix is given by the following approximate expression,
\begin{eqnarray}
	M_\nu\simeq M_{LL}^\dagger-M_{LR}M_{RR}^{-1}M_{LR}^T,
		\label{eq:neutrino_masses}
\end{eqnarray}
where
\begin{eqnarray}
M_{LR}=(G k_1+H k_2e^{-i\alpha})/\sqrt{2},~~~M_{LL}=Fv_Le^{i\theta_L}/\sqrt{2},~~~M_{RR}=Fv_R/\sqrt{2},
\label{eq:neutrino_masses_2}
\end{eqnarray}
with $F$, $G$ and $H$ being $3\times 3$ Yukawa matrices (see Eq.~(\ref{eq:yuk})).
Phenomenologically, the elements of $M_\nu$ must be at the eV scale (or smaller).
The second term in Eq.~(\ref{eq:neutrino_masses}) is the usual ``seesaw'' term -- it
scales roughly as $k_1^2/v_R$, thereby suppressing the neutrinos' masses.
If $v_R$ is only ``moderately'' large (of order 10's of TeV), 
the Yukawa matrices $G$ and $H$ need to be suppressed in order that the elements in this term are
at the eV scale.\footnote{Such a suppression can be obtained in the horizontal symmetry scheme that we
adopt in this paper~\cite{khasanov}, although we do not consider neutrino masses further here.}
The first term in Eq.~(\ref{eq:neutrino_masses}) also benefits from a seesaw relation in that it is
proportional to $v_L\sim k_1^2/v_R$.  The seesaw suppression
is not enough if $v_R={\cal O}(10~\mbox{TeV})$, however,
since then $v_L$ is of order a few GeV (as noted above).
To solve the problem, one could suppress $M_{LL}$ by forcing
the elements of $F$ to be very tiny (of order $10^{-10}$, say), but then the right-handed 
Majorana mass terms
($M_{RR}$) would also become very small (of order keV for $v_R$ of order 10's of TeV).  This
would render the seesaw mechanism for neutrino masses inoperative 
(in fact, Eq.~(\ref{eq:neutrino_masses}) would
become invalid) and would generically yield
GeV-scale neutrinos.  To remedy the situation one needs to force the coefficients $\beta_i$ 
in Eq.~(\ref{eq:A5}) to
be very small or zero, thereby forcing $v_L$ to be very small or zero.  
The $\beta_i$ can be made to disappear
via an exact symmetry, but, as shown in Ref.~\cite{deshpande}, 
the disappearance of the $\beta_i$ comes at the expense of Majorana or Dirac mass terms in this case.

A simple approach that suppresses the $\beta_i$ without eliminating them completely
is to adopt an approximate horizontal $U(1)$ symmetry that is broken by a 
small parameter $\epsilon_H$~\cite{froggatt, nir1993,khasanov}.  In this approach, each of the fields
in the theory is assigned a charge under the symmetry, and terms
in the Lagrangian that are not singlets under this symmetry become 
suppressed by $\epsilon_H^n$, where $n$ is some
combination of the charges of the fields involved.  We adopt the charge assignments~\cite{khasanov}
\begin{eqnarray}
	Q(\Delta_L)&=&-Q(\Delta_R)=-8 , \\
	Q(\phi)&=& -2 ,
	\label{eq:charges}
\end{eqnarray}
with the result that the dimensionless coefficients in the Higgs potential scale as follows
\begin{eqnarray}
	\begin{array}{rcl}
	\alpha_1,\alpha_3 \sim {\cal O}(1),~~~
	\alpha_2 & \sim & {\cal O}(\epsilon_H^4) \sim {\cal O}(\xi) \\
	\beta_1 \sim {\cal O}(\epsilon_H^{16}),~~~
	\beta_2 & \sim & {\cal O}(\epsilon_H^{20}),~~~
	\beta_3 \sim {\cal O}(\epsilon_H^{12})\\
	\lambda_1,\lambda_3 \sim {\cal O}(1),~~~
	\lambda_2 \sim {\cal O}(\epsilon_H^8) & \sim & {\cal O}(\xi^2),~~~
	\lambda_4 \sim {\cal O}(\epsilon_H^4) \sim {\cal O}(\xi)\\
	\rho_1,\rho_2, \rho_3  \sim  {\cal O}(1), && \rho_4\sim{\cal O}(\epsilon_H^{32}) \; .\\
	\end{array}\label{eq:horizontal_scaling}
\end{eqnarray}	
In the numerical work below we set 
$\epsilon_H=0.3$ and $\xi=k_2/k_1= 3/181\simeq 0.017$~\cite{kiersLR}, so
that $\epsilon_H^4\sim\xi$, as noted in the expressions above.  
Furthermore, for a moderate right-handed scale
(of order 15~TeV), $\epsilon=k_1/v_R\sim (246~\mbox{GeV}/(1.5\times 10^{4}~\mbox{GeV})\sim 0.016$,
so that $\epsilon\sim\xi$.  (Of course, for very large right-handed
scales $\epsilon$ becomes much less than $\xi$.)
The $\beta_i$ are particularly suppressed in this scheme, with the result that the
VEV seesaw relation now becomes
\begin{eqnarray}
	v_L\sim \epsilon_H^{20} \epsilon^2 v_R .
		\label{eq:vLapprox}
\end{eqnarray}
Taking $v_R={\cal O}(15$~TeV$)$, we have $v_L\sim 0.1$~eV, which is phenomenologically acceptable.

\subsection{Approximate Expressions for the Minimization Equations}
\label{sec:subsecII.C}

It is useful to consider approximate versions of the six minimization equations 
(Eqs.~(\ref{eq:mu1full})-(\ref{eq:A7})), which can be
obtained by expanding in the small parameters $\xi$, $\epsilon$ and $v_L/v_R$
(as well as taking into account the charge assignments for the various Higgs
potential coefficients).  
Of these, we consider $\epsilon$
and $\xi$ to be of approximately the same order (approximately $0.017$), while $v_L/v_R$ 
is miniscule (of order $10^{-14}$).
We also bear in mind the scaling of the various coefficients in (\ref{eq:horizontal_scaling}), 
recalling that $\epsilon_H^4\sim\xi$.
An approximate (and slightly rearranged) version of Eq.~(\ref{eq:A7})
has already been given above in Eq.~(\ref{eq:A7approx}).
(The corrections to this expression are of order $\epsilon^2\xi$ and $\xi^3$.)  
Furthermore, Eq.~(\ref{eq:A6}) does not need to be expanded at all, since
each of the three terms is of approximately the same order.
From the other four equations we obtain the following expressions,
\begin{eqnarray}
	\frac{\mu_1^2}{v_R^2} & = & \frac{\alpha_1}{2}-
		\frac{1}{2}\alpha_3\xi^2 + 
		\lambda_1\epsilon^2 + {\cal O}(\xi^4,\xi^2\epsilon^2) \; ,
		\label{eq:mu1approx}\\
	\frac{\mu_2^2}{v_R^2} & = & \frac{\alpha_2}{2}
		\frac{\cos(\alpha+\delta_2)}{\cos\alpha}+
		\frac{\alpha_3\xi}{4\cos\alpha} + 
			{\cal O}(\xi^3,\xi\epsilon^2) \; ,
		\label{eq:mu2approx}\\
	\frac{\mu_3^2}{v_R^2} & = & \rho_1 + 
		\frac{1}{2}\alpha_1\epsilon^2 +
		{\cal O}(\xi^2\epsilon^2) ,
		\label{eq:mu3approx}\\
	v_L &=& \frac{\left[\beta_1\xi\cos(\theta_L-\alpha)+\beta_2\cos\theta_L
		+\beta_3\xi^2\cos(\theta_L-2\alpha)\right]\epsilon^2v_R}
			{(2\rho_1-\rho_3)}\left[1+{\cal O}(\xi^2\epsilon^2)\right] \; .
		\label{eq:A5approx}
\end{eqnarray}

Let us first consider the decoupling limit for the Higgs
fields; i.e. let $\epsilon$ become very small,
keeping $\xi$ and $\epsilon_H$ fixed.  Suppose that
we are given a particular model for the Higgs potential, so that
all the coefficients in the Higgs potential are fixed.  In the
decoupling limit,
Eqs.~(\ref{eq:A7approx}) and (\ref{eq:mu1approx})-(\ref{eq:mu3approx}) represent
{\em four} equations in {\em three} unknowns ($v_R$, $\xi$ and $\alpha$).
The system is overconstrained and there must be fine tuning (at order
$\epsilon^2$) among the coefficients in the Higgs potential.  
This fine-tuning is quite severe if $v_R$ is ${\cal O}(10^7~\mbox{GeV})$ 
(as is often assumed in the decoupling limit), in which case 
$\epsilon^2={\cal O}(10^{-9})$.
The remaining two equations (Eqs.~(\ref{eq:A6}) and (\ref{eq:A5approx}))
provide constraints on the remaining three parameters ($v_L$, $\epsilon$
and $\theta_L$).  

It is well-known that the Higgs sector of the
LR Model contains fine-tuning.  In some cases, this fine-tuning extends to 
more than one equation that must be satisfied among coefficients in the
Higgs potential~\cite{bgnr}.  Having said this, we note that the
fine-tuning problem becomes less severe if $v_R$ is 
taken to be at a moderate scale; i.e., of order 15 TeV.  Suppose
again that one is given a set of coefficients in the Higgs potential and that one
wishes to solve the approximate expressions in Eqs.~(\ref{eq:A6}), (\ref{eq:A7approx})
and (\ref{eq:mu1approx})-(\ref{eq:A5approx})
for the six unknowns $v_R$, $v_L$, $\theta_L$, $\epsilon$, $\xi$
and $\alpha$.  One could proceed as follows\footnote{This procedure 
could be made more precise by using the full expressions
for the six first-derivative constraint equations.}:
\begin{enumerate}
\item Use Eq.~(\ref{eq:mu3approx}) (with $\epsilon\approx 0$) to determine $v_R$.
\item Use Eqs.~(\ref{eq:A7approx}) and (\ref{eq:mu2approx}) to
determine $\alpha$ and $\xi=k_2/k_1$.
\item Use Eq.~(\ref{eq:A6}) 
to determine $\theta_L$.
\item Use Eq.~(\ref{eq:mu1approx}) to determine $\epsilon=k_1/v_R$ (thereby
determining the weak scale).
\item Use Eq.~(\ref{eq:A5approx})
to determine $v_L$.
\end{enumerate}
The fine-tuning issue surfaces when one uses Eq.~(\ref{eq:mu1approx}) to determine $\epsilon=k_1/v_R$.
If the dimensionful parameter
$\mu_1^2$ is taken to be ``naturally'' 
of ${\cal O}(v_R^2)$ (as is $\mu_3^2$, from Eq.~(\ref{eq:mu3approx})),
then $\alpha_1$ must equal $2\mu_1^2/v_R^2$ to ${\cal O}(\epsilon^2, \xi^2)\sim 10^{-4}$;
that is, one requires an ${\cal O}(\epsilon^2, \xi^2)$ cancellation between two quantities that
are naturally of order unity.  Alternatively, one could suppress $\mu_1^2$ ``by hand,'' so that it
is of order $k_1^2=\epsilon^2v_R^2$.  In this case $\alpha_1$ also needs to be
suppressed by hand so that it is of order $\epsilon^2$.
In our numerical work
we adopt the latter approach\footnote{At first glance it might
appear that another
source of fine-tuning occurs in Eq.~(\ref{eq:mu2approx}), since the terms on the right-hand
side of the expression are of order $\xi$.  Such is not the case, however, since
$\mu_2^2$ scales like $\epsilon_H^4\sim\xi$, so a ``natural'' choice for $\mu_2^2$ is
$\mu_2^2\sim \xi v_R^2$.}.
 
\section{Higgs Boson Masses}
\label{sec:mass_matrices}

In this section we give approximate expressions for the masses of the various Higgs bosons.  
These expressions help determine phenomenologically
viable scenarios for the Higgs sector of the LR Model.  
The Higgs boson mass matrices are detemined by taking double partial derivatives
of the Higgs potential with respect to the relevant Higgs fields.  The neutral mass
matrix is $8\times 8$, real and symmetric.  The singly- and doubly-charged mass matrices are
both Hermitian and are $4\times 4$ and $2\times 2$, respectively.  The eigenvalues of these
matrices correspond to the squares of the various Higgs boson masses and they must be positive.
Both the neutral and the singly-charged mass matrices have two zero eigenvalues, 
corresponding to the Goldstone modes absorbed to give masses to the $Z$ and $W$ bosons, respectively.

Equations (\ref{eq:mSMapprox})-(\ref{eq:mrho2approx}) below give approximate expressions for the 
squares of the Higgs boson masses. The derivation of these expressions is explained in more 
detail in the Appendix. For the neutral Higgs bosons we obtain the following approximate 
expressions (aside from the two massless Goldstone modes),
\begin{eqnarray}
	m_\textrm{\scriptsize{SM}}^{(0)2} &\simeq& \left[2\lambda_1(1+\xi^2)
		+8\lambda_3\xi^2+8\lambda_4\xi\cos\alpha \right]k_1^2 \; ,
		\label{eq:mSMapprox}\\
	m_{\alpha_3}^{(0)2}&\simeq&\left[\frac{\alpha_3}{2}\frac{(1+\xi^2)}{(1-\xi^2)}
			+2\lambda_3\epsilon^2\right]v_R^2 ~~~~\mbox{(two)} \; , 
		\label{eq:malpha3_0}\\
	m_{\rho_3}^{(0)2}&\simeq&\left(\frac{1}{2}\rho_3-\rho_1+
		4\alpha_2\epsilon^2\xi\sin\alpha\sin\delta_2\right)v_R^2 ~~~~\mbox{(two)} \; , 
		\label{eq:mrho3approx_0}\\
	m_{\rho_1}^{(0)2}&\simeq& 2\rho_1 v_R^2\; ,
		\label{eq:rho1approx}	
\end{eqnarray}
where the word ``two'' in parentheses indicates pairs of degenerate or nearly degenerate eigenvalues.
Note that two of the neutral mass-squared eigenvalues do depend somewhat on $\alpha$, the CP-odd phase
that appears in $\langle\phi\rangle$ (see Eq.~(\ref{eq:HiggsVEVs})). Explicit reference
to the phase $\delta_2$ (the sole CP-odd phase in the Higgs potential) may be eliminated in the 
expression for $m_{\rho_3}^{(0)2}$ by using the first-derivative condition
in Eq.~(\ref{eq:A7}). We have left the expression as is since it is more compact.
For the singly-charged Higgs bosons we obtain the 
following expressions for the two non-Goldstone modes,
\begin{eqnarray}
	m_{\alpha_3}^{(+)2}&\simeq&\frac{\alpha_3}{2}\left[\frac{(1+\xi^2)}{(1-\xi^2)}
		+\frac{1}{2}\epsilon^2(1-\xi^2)\right]v_R^2 \; ,
		\label{eq:malpha3_1}\\
	m_{\rho_3}^{(+)2}&\simeq& m_{\rho_3}^{(0)2} +\frac{\alpha_3}{4}\epsilon^2(1-\xi^2)v_R^2 \; ,
		\label{eq:mrho3approx_1}
\end{eqnarray}
and for the doubly-charged Higgs bosons we have,
\begin{eqnarray}
	m_{\rho_3}^{(++)2}&\simeq&m_{\rho_3}^{(0)2} +\frac{\alpha_3}{2}\epsilon^2(1-\xi^2)v_R^2 \; ,
		\label{eq:mrho3approx_2}\\
	m_{\rho_2}^{(++)2}&\simeq&\left[2\rho_2+\frac{\alpha_3}{2}\epsilon^2(1-\xi^2)\right]v_R^2 \; .
		\label{eq:mrho2approx}
\end{eqnarray}
The expected corrections to the above expressions are described in detail in the Appendix.
At this point we simply note that for the range of parameters considered
in this work the expressions for the singly- and doubly-charged Higgs bosons are essentially
exact, as is Eq.~(\ref{eq:mrho3approx_0}) for two of the neutral Higgs bosons.  
The other expressions for the neutral Higgs bosons
have varying levels of accuracy.  Reference~\cite{gluza} contains approximate expressions
for the Higgs boson masses assuming $v_L=0$ and $\alpha=\delta_2=0$.  The above expressions
agree with those in Ref.~\cite{gluza} in the stated limit, and
neglecting some higher-order corrections.
Also, to a very good approximation, 
$m_{\rho_3}^{(0)2}+m_{\rho_3}^{(++)2}=2m_{\rho_3}^{(+)2}$.\footnote{A
similar relation was reported in Ref.~\cite{datta}, although our expressions
differ in some respects from the ones contained in that paper.  For example,
the authors of Ref.~\cite{datta} include an extra term in the Higgs potential,
and their expressions for the neutral mass matrices appear to be incorrect by an
overall factor of 2.  See also Ref.~\cite{Mohapatrapal1986}.}

A few comments are in order.  First of all, the neutral
Higgs boson with essentially SM-like, flavour-diagonal
couplings to the quarks (except for small corrections)
also generically has a mass that is at the weak
scale.  Furthermore, as was noted above, the six first-derivative
equations allow $\alpha_3$ to be of order unity, so the masses
of the FCNH bosons (Eq.~(\ref{eq:malpha3_0})) are comfortably
of order $v_R$.  In fact, 
in the horizontal symmetry scheme that we
employ, $\lambda_1$, $\alpha_3$, $\rho_1$, $\rho_2$ and $\rho_3$ 
are all of order unity, so all non-SM Higgs bosons have masses that are generically 
of order $v_R$.
This situation is to be contrasted with the case
in which the Higgs potential is CP-invariant ($\sin\delta_2=0$).  In that 
case  it is difficult to maintain a separation in scales between the SM-like Higgs boson
and the non-SM-like Higgs bosons~\cite{bgnr}.

In addition to enforcing the first-derivative conditions (Eqs.~(\ref{eq:mu1full})-(\ref{eq:A7})),
we also need to ensure that the extrema of the potential are actually minima.  Physically, this 
reduces to the requirement that all of the mass-squared eigenvalues be positive.
To a good approximation it is sufficient to require that 
$\lambda_1$, $\alpha_3$, $\rho_1$ and $\rho_2$ all be positive
and that $\rho_3$ be larger than $2\rho_1$.  
The value of the potential at the minimum is approximately $-\rho_1 v_R^4/4$, which is negative,
since $\rho_1$ is positive.

\section{Numerical Results}
\label{sec:numerical}

In this section we perform a numerical analysis of the model to determine
if there are in fact combinations of coefficients in the Higgs potential that
give phenomenologically viable spectra of Higgs bosons.  We also wish to determine whether there might
be correlations among the magnitudes and phases of the various VEVs, or whether
there might be constraints on the sizes of the phases $\alpha$ and $\theta_L$.  Such correlations 
or constraints would affect quark and lepton mixings and
could have important phenomenological consequences in $K$-$\overline{K}$ and $B$-$\overline{B}$ 
mixing and in neutrino physics.

Our basic approach is to select sets of coefficients consistent with
Eqs.~(\ref{eq:mu1full})-(\ref{eq:A7}) and in the ranges specified by 
Eq.~(\ref{eq:horizontal_scaling}) (except for $\alpha_1$, which is taken
to be of order $\epsilon^2$, as noted in Sec.~\ref{sec:subsecII.C}) 
and then to compute the masses of the various Higgs bosons
by diagonalizing the mass matrices numerically.  This procedure also allows
us to check the reliability of the approximate expressions for the masses
in Eqs.~(\ref{eq:mSMapprox})-(\ref{eq:mrho2approx}).  

As noted above,
the mass-squared eigenvalues for physical Higgs bosons must be positive, which
essentially forces $\lambda_1$, $\alpha_3$, $\rho_1$ and $\rho_2$ all to be positive
and $\rho_3$ to be larger than $2\rho_1$.  Let us consider some further
constraints that follow from 
direct and indirect limits on the masses of the Higgs bosons.  

The SM-like Higgs boson has couplings to quarks that are very similar to those
of the Higgs boson in the SM, so we may use limits on the SM Higgs boson
to constrain $\lambda_1$.  The Particle Data Group suggests the 
range $114.4$~GeV$<m_\textrm{\scriptsize{SM}}^{(0)}<193$~GeV for the SM Higgs boson based
on a global fit~\cite{PDG2004}, which translates into the range
$0.1\alt\lambda_1\alt0.3$.
Direct lower bounds from LEP experiments on the singly-charged and doubly-charged Higgs
bosons are typically of order $100$~GeV~\cite{ALEPH2002,OPAL2002,DELPHI2003,L32003}, while
indirect searches can produce a much higher reach.  For example, under certain assumptions 
regarding Yukawa coupling constants the lower bound on doubly-charged Higgs boson masses extends 
to the TeV range~\cite{L32003,OPAL2003}. Future indirect probes of the TeV-range masses are
warranted in other high and low-energy (M\o ller-scattering) experiments.
In the neutral case, the Higgs bosons with masses determined primarily by $\alpha_3$ have flavour
non-diagonal couplings to quarks and can contribute to neutral meson mixing at tree level.  
Limits from such processes are somewhat model dependent, but can provide lower bounds
in the range of several TeV and even up into the range of tens 
of TeV~\cite{ecker85,pospelov,ball1,kiersLR}.
For the purpose of our numerical analysis we fix $v_R=15$~TeV (as well as 
$k_1\simeq 246$~GeV and $\xi=k_2/k_1\simeq 0.0166$) and place lower
bounds on $\alpha_3$, $\rho_1$, $\rho_2$ and $\rho_3$ in such a way that all non-SM 
Higgs bosons obtain masses of order 5~TeV or higher.
For the upper bounds on these coefficients we somewhat arbitrarily choose the 
value ``2,'' which is consistent with ``order unity.''
The bounds employed for all coefficients are described in Table~\ref{table:one}.
The only coefficient that does not scale as in Eq.~(\ref{eq:horizontal_scaling}) is
$\alpha_1$, which we have suppressed ``by hand'' as discussed in Sec.~\ref{sec:subsecII.C}.

\begin{table*}
\caption{Ranges used in the numerical analysis for the dimensionless coefficients in the 
Higgs potential, assuming $v_R=15$~TeV.
The horizontal symmetry breaking parameter $\epsilon_H$
is set to $0.3$ in our numerical work.  All coefficients not included in this table
scale as in Eq.~(\ref{eq:horizontal_scaling}), with minimum and maximum values
of $\mp 2$ multiplied by $\epsilon_H$ to the appropriate power.  Thus,
for example, we require that $-2\epsilon_H^{16}\leq \beta_1\leq 2\epsilon_H^{16}$.}
\label{table:one}
\begin{ruledtabular}
\begin{tabular}{ccc}
	coefficient & minimum value & maximum value \\ \colrule
	$\lambda_1$ & 0.1 & 0.3 \\
	$\alpha_3$ & 0.2 & 2 \\
	$\rho_1$, $\rho_2$ & 0.06 & 2 \\
	$\rho_3$ & $2\rho_1+0.2$ & 2 \\
	$\alpha_1/\epsilon^2$ & $-2$ & 2 \\
	$\alpha_2/\epsilon_H^4$ & 0 & 2 \\
	$\delta_2$ & 0 & $2\pi$ \\
\end{tabular}
\end{ruledtabular}
\end{table*}

Our specific procedure for obtaining VEVs and sets of coefficients that satisfy the 
six constraints in Eqs.~(\ref{eq:mu1full})-(\ref{eq:A7}) is as follows.
First, we choose a random value for $v_L$ (between zero and $2\epsilon_H^{20}k_1^2/v_R$), as
well as random values for the phases $\alpha$ and $\theta_L$ (in the range
zero to $2\pi$).  In a similar manner we choose
random values for $\delta_2$ and all other coefficients in the Higgs potential except for $\alpha_2$, 
$\rho_3$, the
$\mu_i^2$ and one of the $\beta_i$.  The six first-derivative
equations (Eqs.~(\ref{eq:mu1full})-(\ref{eq:A7}))
are then used to compute these remaining six
coefficients.  If $\alpha_2$, $\rho_3$ and the one remaining $\beta_i$ 
are all in acceptable ranges (as determined by Table~\ref{table:one}) we proceed
to compute the various mass-squared eigenvalues.  We do not enforce any particular
constraints for the $\mu_i^2$.  Numerically, $\mu_3^2$ is positive and of order 
$v_R^2$, as one would expect from Eq.~(\ref{eq:mu3approx}) (recall that $\rho_1$ is positive).
$\mu_1^2$ is numerically of order the weak scale squared in magnitude
and $\mu_2^2$ is intermediate between the squares of the weak
and right-handed scales in magnitude.
$\mu_1^2$ and $\mu_2^2$ can be positive or negative, as one might expect
from Eqs.~(\ref{eq:mu1approx}) and (\ref{eq:mu2approx}).  This result may seem counterintuitive
compared to the SM.  In the LR model under consideration it is unusual for the Higgs
potential to be a local maximum at the origin.  The origin is generically a 
generalized saddle point in the sense that the eigenvalues of the second-derivative
matrix generically have mixed signs.  In any case, the Higgs potential is never a local 
minimum at the origin, since $\mu_3^2>0$.\footnote{In the 
eight-dimensional space spanned by the neutral fields
$\phi_1^{0r}$, etc. (with all VEVs and phases set to zero), a local maximum at the origin
is obtained if $\mu_3^2>0$, $2\mu_2^2<\mu_1^2$ and $-2\mu_2^2<\mu_1^2$.  This situation is
``unusual,'' since $\mu_1^2$ is typically much smaller in magnitude than is $\mu_2^2$.
The condition for a local minimum at the origin is obtained by reversing the inequalities
in the three expressions.}

\begin{figure}[t]
\resizebox{5in}{!}{\includegraphics{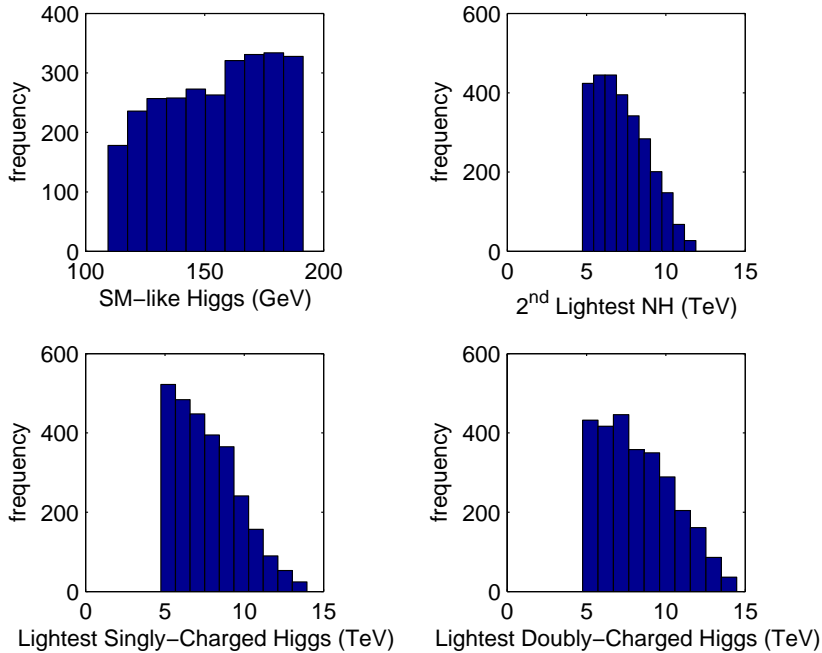}}
\caption{Frequency plot for the masses of the lightest Higgs bosons.  The upper left plot
shows the results for the SM-like Higgs boson, while the other plots show
results for the lightest neutral, singly-charged and doubly-charged nonstandard Higgs bosons.}
\label{fig:mass_hist}
\end{figure}
\begin{figure}[t]
\resizebox{5in}{!}{\includegraphics{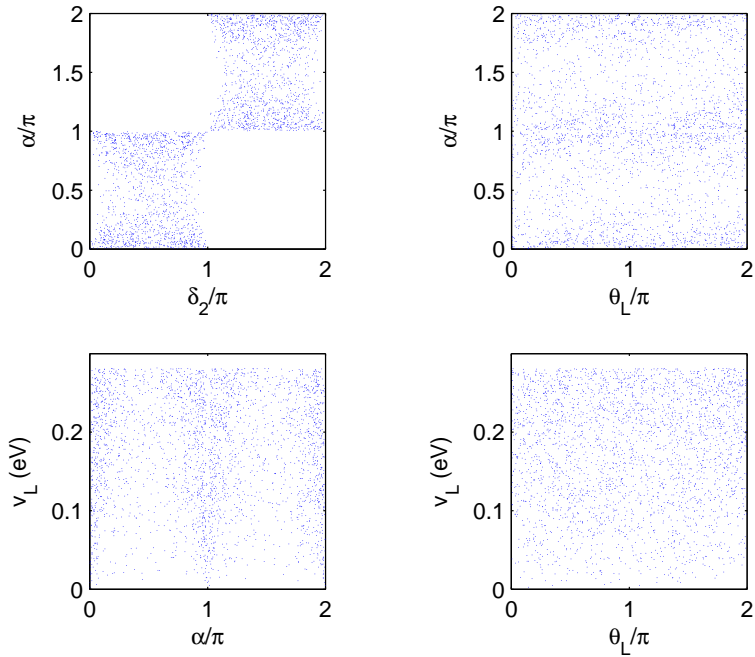}}
\caption{Various correlations between $\alpha$, $\delta_2$, $\theta_L$ and $v_L$.}
\label{fig:phases}
\end{figure}

Figures~\ref{fig:mass_hist} and \ref{fig:phases} show the results of our numerical analysis.
The first of these shows a frequency plot of the lightest Higgs boson masses.
While we have not imposed any direct mass constraints on the numerical results,
the ranges described in Table~\ref{table:one} effectively limit the mass of the SM-like Higgs
boson to be in the range $100~\mbox{GeV}\alt m_\textrm{\scriptsize{SM}}^{(0)} \alt 200$~GeV
and those of all nonstandard Higgs bosons to be of order 4.7~TeV or greater.  The scatter plots
in Fig.~\ref{fig:phases} show various correlations between $v_L$, $\theta_L$, $\alpha$ and $\delta_2$.
The correlation between $\alpha$ and $\delta_2$ is easily understood by examining the
approximate relation in Eq.~(\ref{eq:A7approx}) -- since $\alpha_2$, $\alpha_3$ and $\xi$
are all positive quantities, $\sin\delta_2$ and $\sin\alpha$ must have the same sign.
The remaining three plots show various combinations of $\alpha$, $v_L$ and $\theta_L$, 
quantities that affect
quark and lepton masses and mixings.  In particular, notice that the phases
$\alpha=\arg\left(\langle\phi_2^0\rangle\right)$ and 
$\theta_L=\arg\left(\langle\delta_L^0\rangle\right)$ are not constrained to be small.
Also, $v_L$ is of order $0.1$~eV, which is phenomenologically viable.
From these plots it does not appear that there are
strong correlations between $\alpha$, $v_L$ and $\theta_L$.
A numerical investigation of the case $v_R=50$~TeV yields similar plots (except that
$v_L$ is reduced in magnitude because it scales as $1/v_R$).

Our numerical analysis also allows us to check the reliability of the approximate expressions 
for the mass-squared eigenvalues in
Eqs.~(\ref{eq:mSMapprox})-(\ref{eq:mrho2approx}).
The eigenvalues were computed numerically from the mass
matrices and compared to the approximate expressions.  
For the singly- and doubly-charged Higgs bosons,
the ``exact'' (numerical) values agreed with the approximate
expressions to within a few parts in $10^{15}$ for $v_R=15$~TeV.
The level of agreement for $m_{\rho_3}^{(0)2}$ was similar.
Corrections for the other four neutral mass-squared eigenvalue expressions
are described in the Appendix.

\section{Discussion and Conclusions}
\label{sec:conclusions}

We have considered the Higgs sector of the minimal Left-Right model with explicit CP-violation in 
the Higgs potential in the ``moderate'' decoupling limit, i.e., when the scale set by the 
VEV of the right-handed Higgs field $v_R$ is in the range 15-50 TeV. This intermediate regime 
provides (at least in principle) testable effects of the remnants of RH symmetries in
upcoming collider experiments. At the same time, inclusion of explicit CP violation 
in the Higgs potential allows for the generation of a viable spectrum of Higgs bosons containing 
one SM-like Higgs boson with mass of order the weak scale and several heavy 
neutral and charged Higgs bosons with masses of order $v_R$. Supplemented 
by an additional $U(1)$ horizontal symmetry, this model alleviates most of the fine-tuning
effects associated with minimal LR models in the decoupling regime~\cite{bgnr}, resulting in
an SM-like low energy effective theory. Yet, some small amount of fine-tuning is still 
required. We have also performed numerical simulations of 
the Higgs spectrum in our model, confirming our phenomenological analysis and the
reliability of the adopted approximations. The power counting for the Higgs potential 
coupling constants provided by the broken $U(1)$ charge assignments also allows 
for the natural generation of neutrino masses in such a model.

\begin{acknowledgments}
It is a pleasure to thank A.~Soni and G.-H.~Wu for helpful conversations and for 
collaboration at an early stage of this work. The work of K.K.and M.A.~was supported in part by the 
U.S.\ National Science Foundation under Grant PHY--0301964.
A.P.~was supported in part by the U.S.\ National Science Foundation under Grant 
PHY--0244853, and by the U.S.\ Department of Energy under Contract DE-FG02-96ER41005.
\end{acknowledgments}

\appendix*

\section{Goldstone Bosons and Approximate Masses}
\label{appendix}

In this appendix we give exact expressions for the charged and neutral Goldstone bosons.
These expressions are used to construct matrices that perform exact block diagonalizations
for the singly-charged and neutral mass matrices, separating out
the ``zero'' eigenvalues for the Goldstone modes.  We also explain our procedure for obtaining
the approximate mass-squared eigenvalues given in the text 
(Eqs.~(\ref{eq:mSMapprox})-(\ref{eq:mrho2approx})).

\subsection{Goldstone Bosons}

The kinetic terms for the Higgs boson fields in the Lagrangian are given by~\cite{duka}
\begin{eqnarray}
{\cal L}_\textrm{\scriptsize{kin}}= \mbox{Tr}\left[\left(D_\mu\phi\right)^\dagger D^\mu\phi\right]
		+\mbox{Tr}\left[\left(D_\mu\Delta_L\right)^\dagger D^\mu\Delta_L\right]
		+\mbox{Tr}\left[\left(D_\mu\Delta_R\right)^\dagger D^\mu\Delta_R\right],
\label{eq:kin_Higgs}
\end{eqnarray}
where covariant derivatives are defined as
\begin{eqnarray}
D_\mu\phi&=&\partial_\mu\phi-\frac{ig}{2}W_{L\mu}^a\tau^a\phi+\frac{ig}{2}\phi W_{R\mu}^a\tau^a \; ,\\
D_\mu\Delta_{L,R}&=&\partial_\mu\Delta_{L,R}-\frac{ig}{2}\left[W_{L,R\mu}^a\tau^a,\Delta_{L,R}\right]
		-ig^\prime B_\mu\Delta_{L,R} \; .
\end{eqnarray}
Inserting Eqs.~(\ref{eq:PhiDeltaLR}) and (\ref{eq:phi10})-(\ref{eq:deltaR0}) into 
Eq.~(\ref{eq:kin_Higgs}) yields the mass matrices for the gauge
bosons, which may be diagonalized to obtain the physical mass eigenstates 
($W_{1,2}^{\pm\mu}$, $Z_{1,2}$ and $A_\mu$) in terms of the original gauge bosons 
associated with $SU(2)_L\times SU(2)_R\times U(1)_{B-L}$ 
($W_{L,R}^{\pm\mu}=\left(W_{L,R}^{1\mu}\mp i W_{L,R}^{2\mu}\right)/\sqrt{2}$, 
$W_{L,R}^{3\mu}$ and $B_\mu$). Equation~(\ref{eq:kin_Higgs}) also yields bilinear 
couplings between the physical gauge bosons and various linear
combinations of the Higgs fields.  One such term is proportional to
$ig W_{1\mu}^-\partial^\mu G_1^+$, for example, and allows one to identify $G_1^+$ as a (would-be) 
Goldstone boson.  Proceeding in this manner one obtains expressions for the two charged, orthogonalized
Goldstone bosons,
\begin{eqnarray}
G_1^+&=&\frac{1}{N_1^+}\left[\left(k_1\sin\zeta-k_2\cos\zeta\right)e^{-i\alpha}\phi_1^+
		+\left(k_1\cos\zeta-k_2\sin\zeta\right)\phi_2^+\right. \nonumber\\
		&& \left. ~~~~~-\sqrt{2}v_Re^{-i\alpha}\sin\zeta \delta_R^+
				-\sqrt{2}v_Le^{-i\theta_L}\cos\zeta\delta_L^+\right] , \\
G_2^+&=&\frac{1}{N_2^+}\left[\left(k_1\cos\zeta+k_2\sin\zeta\right)e^{-i\alpha}\phi_1^+
		+\left(-k_1\sin\zeta-k_2\cos\zeta\right)\phi_2^+\right. \nonumber\\
		&& \left. ~~~~~-\sqrt{2}v_Re^{-i\alpha}\cos\zeta \delta_R^+
				+\sqrt{2}v_Le^{-i\theta_L}\sin\zeta\delta_L^+\right] ,
\end{eqnarray}
where $N_{1,2}^+$ are normalization constants and
\begin{eqnarray}
\sin 2\zeta = \frac{2k_1k_2}{\left[(v_R^2-v_L^2)^2+4k_1^2k_2^2\right]^{1/2}} \; .
\end{eqnarray}
We may also construct two other normalized, orthogonal linear combinations of the four fields $\phi_1^+$,
$\phi_2^+$, $\delta_R^+$ and $\delta_L^+$,
\begin{eqnarray}
\phi_a^+&=&\frac{1}{N_a^+}\left(e^{-i\alpha}\phi_1^++r_{32}^+\phi_2^+
	+r_{33}^+e^{-i\alpha}\delta_R^+\right) \; ,\\
\phi_b^+&=&\frac{1}{N_b^+}\left(r_{41}^+e^{-i\alpha}\phi_1^++r_{42}^+\phi_2^+
	+r_{43}^+e^{-i\alpha}\delta_R^++e^{-i\theta_L}\delta_L^+\right) \; ,
\end{eqnarray}
where the $r_{ij}^+$ are simple functions of $k_1$, $k_2$, $v_R$ and $v_L$.
(Forcing $\phi_a^+$ to be orthogonal to $G_1^+$ and $G_2^+$ determines $r_{32}^+$ and $r_{33}^+$;
similarly, forcing $\phi_b^+$ to be orthogonal to $G_1^+$, $G_2^+$ and $\phi_a^+$
determines $r_{41}^+$, $r_{42}^+$ and $r_{43}^+$.)
Putting the above expressions together, we form the unitary matrix ${\cal R}^{(+)}$ that 
relates the two bases,
\begin{eqnarray}
\left(\begin{array}{c}
	G_1^+ \\ G_2^+ \\ \phi_a^+ \\ \phi_b^+ \\
		\end{array}\right) = {\cal R}^{(+)}
\left(\begin{array}{c}
	\phi_1^+ \\ \phi_2^+ \\ \delta_R^+ \\ \delta_L^+ \\
	\end{array}\right) \; .
\end{eqnarray}
The matrix ${\cal R}^{(+)}$ may be used to bring the $4\times 4$ singly-charged mass matrix 
into block-diagonal form, with a non-zero $2\times 2$ block in the lower-right,
\begin{eqnarray}
{\cal R}^{(+)*}{\cal M}_+^2{\cal R}^{(+)T} = 
	\left(\begin{array}{cc}
		0 & 0 \\
		0 & {\cal M}_{+,2\times 2}^2
		\end{array}\right) \; .
\label{eq:Rplusrotn}
\end{eqnarray}

One can similarly block-diagonalize the neutral mass matrix, although the diagonalization
of the neutral gauge bosons is somewhat more involved than that of the charged gauge bosons.
Furthermore, both $g$ and $g^\prime$ are involved, so that the physical mass eigenstates
depend on the Weinberg angle ($\tan^2\theta_W = g^{\prime 2}/\left(g^2+g^{\prime 2}\right)$).
Since our purpose is to block-diagonalize the neutral Higgs sector, and since the Higgs sector
has no dependence on $\theta_W$, we will consider the (unphysical) limit $g^\prime/g\to 0$.  This means
our expressions for the Goldstone modes, while useful for diagonalization purposes, will not 
correspond to the actual combinations of Higgs bosons ``eaten'' by $Z_{1,2}$.
Taking this limit, and proceeding as in the charged Goldstone case, we obtain
\begin{eqnarray}
	G_1^0&=&\frac{1}{N_1^0}\left[\left(\cos\phi+\sin\phi\right)\left(k_1\phi_1^{0i}
		-k_2\phi_2^{0i}\right)+2v_R\sin\phi \delta_R^{0i}
				-2v_L\cos\phi\delta_L^{0i}\right] , \\
	G_2^0&=&\frac{1}{N_2^0}\left[\left(\cos\phi-\sin\phi\right)\left(k_1\phi_1^{0i}
		-k_2\phi_2^{0i}\right)+2v_R\cos\phi \delta_R^{0i}
				+2v_L\sin\phi\delta_L^{0i}\right] ,
\end{eqnarray}
where 
\begin{eqnarray}
	\sin 2\phi = -\frac{k_1^2+k_2^2}{\left[4(v_R^2-v_L^2)^2+(k_1^2+k_2^2)^2\right]^{1/2}} \; .
\end{eqnarray}
Defining orthogonal combinations
\begin{eqnarray}
	\phi_a^{0i}&=&\frac{1}{N_a^0}\left(r_{31}^0\phi_1^{0i}+\phi_2^{0i}\right) =
		\frac{1}{\sqrt{1+\xi^2}}\left(\xi\phi_1^{0i}+\phi_2^{0i}\right)\; ,
		\label{eq:phia0i}\\
	\phi_b^{0i}&=&\frac{1}{N_b^0}\left(r_{41}^0\phi_1^{0i}+r_{42}^0\phi_2^{0i}
		+r_{43}^0\delta_R^{0i}+\delta_L^{0i}\right) 
\end{eqnarray}
(where the $r_{ij}^0$ are again determined by forcing the various fields
to be orthogonal), we construct the orthogonal matrix ${\cal R}^{(0)}$ that relates the two bases,
\begin{eqnarray}
	\left(\begin{array}{c}
		G_1^0 \\ G_2^0 \\ \phi_a^{0i} \\ \phi_b^{0i} \\
		\end{array}\right) = {\cal R}^{(0)}
	\left(\begin{array}{c}
		\phi_1^{0i} \\ \phi_2^{0i} \\ \delta_R^{0i} \\ \delta_L^{0i} \\
		\end{array}\right) \; .
\end{eqnarray}
Defining an $8\times 8$ orthogonal matrix
\begin{eqnarray}
	\widetilde{\cal R}^{(0)} = \left(\begin{array}{cc}
				0 & {\cal R}^{(0)} \\
				\mathbf{1}_{4\times 4} & 0 \\
					\end{array}\right),
\end{eqnarray}
we then have
\begin{eqnarray}
	 \widetilde{\cal R}^{(0)}{\cal M}_0^2\widetilde{\cal R}^{(0)T} = 
		\left(\begin{array}{cc}
			0 & 0 \\
			0 & {\cal M}_{0,6\times 6}^2
			\end{array}\right) \; ,
		\label{eq:R0rotn}
\end{eqnarray}
where ${\cal M}_{0,6\times 6}^2$ is the symmetric $6\times 6$ mass matrix in the basis
$\Phi_0^T=(\phi_a^{0i},\phi_b^{0i},\phi_1^{0r},\phi_2^{0r},\delta_R^{0r},\delta_L^{0r})$. 

\subsection{Approximate Mass-Squared Eigenvalues for Physical Higgs Bosons}

Throughout this paper we assume that $v_L$ is small -- of order $0.1$~eV or less -- so that
it is at a natural scale for neutrino masses. In this paper we also assume that the 
$\beta_i$ are suppressed through a horizontal symmetry, as indicated in 
Eq.~(\ref{eq:horizontal_scaling}), although any other scheme that introduces
numerical scaling similar to Eq.~(\ref{eq:horizontal_scaling}) is indeed acceptable.
Such a suppression of the $\beta_i$ leads naturally to the assumed small value for $v_L$, 
even if $v_R$ is only ``moderately large'' ($v_R\sim 15$~TeV),
as indicated in Eqs.~(\ref{eq:vLapprox}) and (\ref{eq:A5approx}).
It is important in our analysis that $v_L$ and the $\beta_i$ are not identically
zero, since $v_L$ and its associated phase, $\theta_L$, are significant parameters for neutrino physics
(see Eqs.~(\ref{eq:neutrino_masses}) and (\ref{eq:neutrino_masses_2})).
Nevertheless, for calculating Higgs boson masses, it is an excellent approximation to consider the limit
$\beta_i, v_L\to 0$.  This statement will be quantified in the following.

\subsubsection{Doubly-Charged Higgs Bosons}

The doubly-charged Higgs boson masses are determined by a $2\times  2$ Hermitian matrix.
The off-diagonal elements in this matrix are proportional to $\rho_4 v_L v_R$, $\beta_1 k_1 k_2$,
$\beta_2 k_2^2$ and $\beta_3 k_1^2$.  Assuming the scaling given in 
Eq.~(\ref{eq:horizontal_scaling}), the largest of these terms scales approximately as 
${\cal O}(\epsilon^2\epsilon_H^{12}v_R^2)$.
The diagonal elements are of order $v_R^2$.  Assuming no extreme
accidental degeneracies in
the diagonal elements, the corrections to the mass-squared eigenvalues due to the off-diagonal
contributions are of order $\epsilon^4\epsilon_H^{24}v_R^2$.  With $\epsilon\sim 0.016$
(for $v_R\sim 15$~TeV) and $\epsilon_H=0.3$,
$\epsilon^4\epsilon_H^{24}\sim {\cal O}(10^{-20})$, and one can safely take the
limit $\beta_i, v_L\to 0$ for the off-diagonal elements.  Taking this limit for the diagonal elements
is also a very good approximation, since it only
introduces a tiny error (of order $\epsilon^4\epsilon_H^{40} v_R^2\sim10^{-28}\times v_R^2$).  
The approximate eigenvalues
are thus simply the diagonal elements of the doubly-charged mass matrix in the limit
$\beta_i, v_L\to 0$.  
These approximate eigenvalues are denoted by $m_{\rho_3}^{(++)2}$ and $m_{\rho_2}^{(++)2}$
in Eqs.~(\ref{eq:mrho3approx_2}) and (\ref{eq:mrho2approx}) and
correspond to the fields $\delta_L^{++}$ and $\delta_R^{++}$, respectively.
Our numerical work yields excellent agreement between these approximate expressions 
and the values obtained numerically by diagonalizing the full mass matrix -- the
values agree to within a few parts in $10^{15}$ in our numerical study with $v_R=15$~TeV.

\subsubsection{Singly-Charged Higgs Bosons}

For the singly-charged Higgs bosons it is also an excellent approximation to consider
the limit $\beta_i, v_L\to 0$.  The justification for this approximation is somewhat more
intricate to argue than in the doubly-charged
case, but numerically it does appear to be an excellent approximation.  
(The numerical agreement for $v_R=15$~TeV is similar to that found in the doubly-charged case.)
Taking the limit $\beta_i, v_L\to 0$ in the mass matrix and in ${\cal R}^{(+)}$, one 
finds that $\delta_L^+$ decouples from the other three singly-charged Higgs fields,
yielding
\begin{eqnarray}
{\cal M}_{+}^2 \simeq \left(
	\begin{array}{cccc}
	m_+^2 & m_+^2\xi e^{-i\alpha} & \frac{1}{\sqrt{2}}m_+^2 \epsilon(1-\xi^2) & 0 \\
	& & & \\
	m_+^2\xi e^{i\alpha} & m_+^2\xi^2 & \frac{1}{\sqrt{2}}m_+^2\epsilon\xi(1-\xi^2)e^{i\alpha} & 0 \\
	& & & \\
\frac{1}{\sqrt{2}}m_+^2 \epsilon(1-\xi^2) & \frac{1}{\sqrt{2}}m_+^2\epsilon\xi(1-\xi^2)e^{-i\alpha} &
	\frac{1}{2}m_+^2\epsilon^2(1-\xi^2)^2 & 0 \\
	& & & \\
	0 & 0 & 0 & m_{\rho_3}^{(+)^2} \\
	\end{array}\right),
\end{eqnarray}
in the basis $(\phi_1^+,\phi_2^+,\delta_R^+,\delta_L^+)$, where
\begin{eqnarray}
	m_+^2 = \frac{\alpha_3}{2}\frac{v_R^2}{\left(1-\xi^2\right)}
\end{eqnarray}
and where the approximate eigenvalue $m_{\rho_3}^{(+)^2}$ is given in Eq.~(\ref{eq:mrho3approx_1}).
Performing the unitary transformation in Eq.~(\ref{eq:Rplusrotn}) yields
the approximate expression for the remaining non-zero eigenvalue, denoted
$m_{\alpha_3}^{(+)^2}$ in Eq.~(\ref{eq:malpha3_1}).
This latter eigenvalue is associated with the field 
$\phi_a^+\simeq\left[e^{-i\alpha}\phi_1^++\xi\phi_2^+
	+\epsilon(1-\xi^2)e^{-i\alpha}\delta_R^+/\sqrt{2}\right]/N_a^+$.

\subsubsection{Neutral Higgs Bosons}

Calculation of the approximate neutral mass-squared eigenvalues is simplified by once again
taking the limit $\beta_i, v_L\to 0$.  We begin with the 
orthogonal rotation shown in Eq.~(\ref{eq:R0rotn}) (taking the limit
$v_L\to 0$ in $\widetilde{\cal R}^{(0)}$).  This rotation yields the $6 \times 6$ matrix
${\cal M}_{0,6\times 6}^2$, whose eigenvalues
correspond to the six non-Goldstone Higgs bosons.
In the limit $\beta_i, v_L\to 0$, two of the fields 
($\delta_L^{0r}$ and $\phi_b^{0i}\simeq\delta_L^{0i}$)
decouple from the rest, allowing for immediate identification
of their mass-squared eigenvalues.  The eigenvalues are degenerate in this limit and are denoted
by $m_{\rho_3}^{(0)^2}$ in
Eq.~(\ref{eq:mrho3approx_0}).  In numerical tests with $v_R=15$~TeV, the approximate expression for these
eigenvalues has an accuracy comparable to those
for the singly- and doubly-charged masses.
Removing the entries corresponding to these two fields
reduces the size of the remaining mass matrix to $4\times 4$.  We call this matrix
$\widetilde{\cal M}_{0,4\times 4}^2$ in the $(\phi_a^{0i},\phi_1^{0r},\phi_2^{0r},\delta_R^{0r})$ basis,
where $\phi_a^{0i}$ is defined in Eq.~(\ref{eq:phia0i}).
The procedure so far can be illustrated schematically as follows,
\begin{eqnarray}
	{\cal M}_0^2 ~~~ \stackrel{\beta_i, v_L\to 0; \widetilde{\cal R}^{(0)}}
		{\Longrightarrow}~~~{\cal M}_{0,6\times 6}^2
		~~~\stackrel{\textrm{\scriptsize{remove~}}\delta_L^{0r}, \delta_L^{0i}}
		{\Longrightarrow}~~~\widetilde{\cal M}_{0,4\times 4}^2 \; .
\end{eqnarray}
The fields $\phi_a^{0i}$ and $\delta_R^{0r}$ are approximately mass eigenstates, with
the remaining two approximate mass eigenstates being the following,
\begin{eqnarray}
\phi_\textrm{\scriptsize{SM}} & = &\frac{1}{\sqrt{1+\xi^2}}\left(\phi_1^{0r}+\xi\phi_2^{0r}\right)\; , 
		\label{eq:phiSM}\\
\phi_a^{0r} & = & \frac{1}{\sqrt{1+\xi^2}}\left(-\xi\phi_1^{0r}+\phi_2^{0r}\right)\; .
\label{eq:phia}
\end{eqnarray}
Defining
\begin{eqnarray}
	 {\cal R}^{(0)}_2 = 
		\left(\begin{array}{cccc}
			0 & \frac{1}{\sqrt{1+\xi^2}} & \frac{\xi}{\sqrt{1+\xi^2}} & 0 \\
			1 & 0 & 0 & 0 \\
			0 & -\frac{\xi}{\sqrt{1+\xi^2}} & \frac{1}{\sqrt{1+\xi^2}} & 0 \\
			0 & 0 & 0 & 1 \\
			\end{array}\right) \; ,
		\label{eq:R02rotn}
\end{eqnarray}
the mass matrix in the basis 
$(\phi_\textrm{\scriptsize{SM}},\phi_a^{0i},\phi_a^{0r},\delta_R^{0r})$ is
\begin{eqnarray}
	{\cal M}_{0,4\times 4}^2 = {\cal R}^{(0)}_2\widetilde{\cal M}_{0,4\times 4}^2{\cal R}^{(0)T}_2
		\sim \left(\begin{array}{cccc}
			\epsilon^2 & \epsilon^2\xi & \epsilon^2\xi & \epsilon\alpha_1,\epsilon\xi^2 \\
			\epsilon^2\xi  & 1 & \epsilon^2\xi^2 & \epsilon\xi \\
			\epsilon^2\xi & \epsilon^2\xi^2 & 1 & \epsilon\xi \\
			\epsilon\alpha_1,\epsilon\xi^2 & \epsilon\xi & \epsilon\xi & 1\\
			\end{array}\right) v_R^2 \; ,
\end{eqnarray}
where the expression on the right gives the orders of magnitude of each of the entries, taking
into account the scaling in Eq.~(\ref{eq:horizontal_scaling}).  In the 1-4 and 4-1 entries
we have explicitly included a term of order $\epsilon\alpha_1$, which could be of order $\epsilon$ if one
chooses to take $\alpha_1\sim{\cal O}(1)$, but would be similar to other terms in that entry if one
would choose $\alpha_1\sim{\cal O}(\xi^2)$ or ${\cal O}(\epsilon^2)$.  In the following we assume
that $\alpha_1\sim{\cal O}(\epsilon^2)$.\footnote{Upon diagonalization, the ``$\alpha_1$'' term in
the 1-4 element in the mass matrix
makes a contribution of order $\alpha_1^2\epsilon^2v_R^2$
to the SM-like mass-squared eigenvalue $m_\textrm{\scriptsize{SM}}^{(0)2}$.
If $\alpha_1\sim{\cal O}(\epsilon^2)$, this term is negligibly small and may be ignored.
If $\alpha_1\sim{\cal O}(1)$, however, the contribution is similar in magnitude
to the leading contribution ($\sim 2\lambda_1\epsilon^2v_R^2$ -- see Eq.~(\ref{eq:mSMapprox}))
and it must be taken into account carefully.}

The diagonal elements of ${\cal M}_{0,4\times 4}^2$ provide good estimates of the eigenvalues
of the matrix.  One could in principle compute corrections to 
these estimates by assuming that ${\cal M}_{0,4\times 4}^2$ is diagonalized by 
a ``small'' unitary rotation (i.e., by a unitary matrix that is
essentially unity along the diagonal and that has small off-diagonal elements).
Such an approach is complicated by the fact that the
2-2 and 3-3 elements in ${\cal M}_{0,4\times 4}^2$ are nearly degenerate.
In fact, the difference between the two elements is of order $\epsilon^2\xi^2v_R^2$, 
which is the same order of magnitude as 
the 2-3 element of ${\cal M}_{0,4\times 4}^2$.  This situation can lead to a large amount of mixing
in the 2-3 block upon diagonalization of ${\cal M}_{0,4\times 4}^2$ and can also
complicate matters somewhat for computing corrections to the 1-1 and 4-4 elements.
We choose instead simply to use the diagonal elements to estimate the eigenvalues.

The approximate eigenvalue denoted $m_\textrm{\scriptsize{SM}}^{(0)2}$ in Eq.~(\ref{eq:mSMapprox})
corresponds to the field $\phi_\textrm{\scriptsize{SM}}$ in Eq.~(\ref{eq:phiSM}).
Corrections to Eq.~(\ref{eq:phiSM}) are expected to be of order $\epsilon^6v_R^2$,
$\epsilon^4\xi^2v_R^2$ and $\epsilon^2\xi^4v_R^2$, where we have assumed that the
various Higgs potential coefficients scale as in Eq.~(\ref{eq:horizontal_scaling})
(except for $\alpha_1$, which is taken to be of order $\epsilon^2$).
The approximate eigenvalue denoted $m_{\alpha_3}^{(0)2}$
in Eq.~(\ref{eq:malpha3_0}) corresponds to the 
nearly-degenerate fields $\phi_a^{0i}$ and $\phi_a^{0r}$.  Corrections to
Eq.~(\ref{eq:malpha3_0}) are expected to be of order $\epsilon^2\xi^2v_R^2$.
Finally, the approximate eigenvalue corresponding to $\delta_R^{0r}$
is denoted by $m_{\rho_1}^{(0)2}$ in Eq.~(\ref{eq:rho1approx}).  Corrections
to this expression are also expected to be of order $\epsilon^2\xi^2v_R^2$.
In our numerical study with $v_R=15$~TeV we have compared the exact and approximate expressions
for the various mass-squared eigenvalues.  The numerical differences 
are consistent with the quoted corrections
for $m_\textrm{\scriptsize{SM}}^{(0)2}$, $m_{\alpha_3}^{(0)2}$
and $m_{\rho_1}^{(0)2}$, although sometimes the corrections are larger
due to accidental degeneracies or combinations of coefficients that occassionally 
give large enhancements.  An example of the former effect occurs
if the 2-2 and 4-4 or 3-3 and 4-4
elements of ${\cal M}_{0,4\times 4}^2$ are nearly degenerate.  In that case, corrections
to the respective eigenvalues can be of order $\epsilon\xi v_R^2$ instead of
$\epsilon^2\xi^2 v_R^2$.  
An example of the latter effect occurs for $m_\textrm{\scriptsize{SM}}^{(0)2}$,
for which one of the leading corrections is proportional to 
$32\lambda_3^2\epsilon^4\xi^2v_R^2/\alpha_3$.  If $\lambda_3\simeq 2$ and 
$\alpha_3\simeq 0.2$, this correction is of order $640\times \epsilon^4\xi^2v_R^2$.

\newpage


\end{document}